\documentclass[aip,reprint]{revtex4-1}

\usepackage{color}
\usepackage{graphicx}
\usepackage{amsmath}
\usepackage{subcaption}
\usepackage{gensymb,units,textcomp}

\begin{document}
\title{Biermann Battery Magnetic Fields in ICF Capsules: Total Magnetic Flux Generation}

\author{C. A. Walsh}
\email{walsh34@llnl.gov}
%\author{R. C. Nora}
\author{D. S. Clark}
\affiliation{Lawrence Livermore National Laboratory, Livermore, CA 94550 USA}

\date{\today}

	\begin{abstract}
		This paper focuses on the process of magnetic flux generation in ICF implosions. Hot-spots are shown to be dominated by fields generated during stagnation, when the temperature and density gradients are largest. A scaling of hot-spot magnetic flux is derived and compared with simulations, revealing that perturbations with both larger amplitudes and higher mode numbers generate more magnetic flux.
		The model allows for greater understanding of which target designs will be susceptible to MHD effects. For example, the model can be used to ascertain the time when most magnetic flux is generated. If generation is weighted more towards early times, then more high-mode magnetic field loops will be present. A hot-spot with no high-mode perturbations at time of peak neutron production can still contain significant magnetic flux on those scales. By assuming that magnetic flux is deposited at the hot-spot edge by Nernst advection, the model can be used to post-process radiation-hydrodynamics data to estimate magnetic field strengths and magnetizations.
	\end{abstract}
	\maketitle
	
	\section{Introduction}
	Magnetic fields are self-generated in inertial confinement fusion (ICF) implosions by the Biermann Battery mechanism. Previous 3-D extended-magnetohydrodynamic (extended-MHD) Gorgon simulations of an indirect-drive implosion from the NIF high-foot campaign \cite{hurricane2014} demonstrated that these fields could grow to over \unit[10,000]{T} in strength, resulting in magnetized heat-flow (Hall Parameter $>$ 1)\cite{walsh2017}. While thermal conductivities were reduced by over 90\% in some locations, Righi-Leduc heat-flow became an important mechanism for increasing hot-spot heat losses \cite{walsh2017}. Overall, the hot-spot temperature and yield were relatively unaffected by the self-generated magnetic fields. While demonstrating the intrinsic complexity of extended-MHD heat-flow in ICF implosions, the simulations had limitations. Firstly, the simulations did not include magnetic fields generated during the drive-phase of the implosion. Secondly, the capsule implosions simulated did not include native perturbation sources, only applying artificial velocity perturbations once the first shock converged on the axis. It still remains a possibility that magnetized heat-flow could explain discrepancies between experiments and simulations, which do not typically include MHD \cite{clark2016}. More recent simulations show that stagnation-phase perturbation growth is enhanced by self-generated magnetic fields \cite{walsh2021a}.
	
	This paper studies the magnetic flux generation process in more detail using native perturbation sources and simulations over the whole capsule history. Principles of flux generation are demonstrated through a theoretical scaling, which compares favorably to the full extended-MHD simulations. By assuming Nernst deposits all the magnetic flux at the hot-spot edge, the model can predict magnetic field profiles using only the simulated $\rho R$ (line-integrated density), $T_e R$ (line-integrated temperature) and bulk hot-spot quantities such as temperature and density as a function of time. This model can then be used to post-process state of the art radiation hydrodynamics calculations, which are better equipped to resolve perturbation sources relevant to ICF implosions.
	
	The magnetic field $\underline{B}$ in an extended-MHD plasma is governed by the following transport equation \cite{braginskii1965,walsh2020,sadler2021}:
	
	\begin{align}
	\begin{split}
	\frac{\partial \underline{B}}{\partial t} = & - \nabla \times \frac{\eta}{\mu_0 } \nabla \times \underline{B} + \nabla \times (\underline{v}_B \times \underline{B} ) \\
	&+ \nabla \times \Bigg( \frac{\nabla P_e}{e n_e} - \frac{\beta_{\parallel} \nabla T_e}{e}\Bigg) \label{eq:mag_trans_new}
	\end{split}
	\end{align}

	Where the first term on the right is resistive diffusion with diffusivity $\eta$ and the second term is advection of the magnetic field at velocity $\underline{v}_B$:
	
	\begin{equation}
	\label{eq:mag_trans_new_velocity}	\underline{v}_B = \underline{v} - \gamma_{\bot} \nabla T_e - \gamma_{\wedge}(\underline{\hat{b}} \times \nabla T_e) %- \frac{\underline{j}}{e n_e}(1 + \delta_{\bot}^c) + \frac{\delta_{\wedge}^c}{e n_e} (\underline{j} \times \underline{\hat{b}})   
	\end{equation}
	
	Here the current-driven transport terms have been neglected due to their insignificance for Biermann Battery generated magnetic fields \cite{walsh2020}. The total advection velocity $\underline{v}_B$ includes transport with the bulk plasma motion $\underline{v}$, down temperature gradients $-\nabla T_e$ (called the Nernst term) and perpendicular to both the magnetic field and the temperature gradients $-\underline{\hat{b}} \times \nabla T_e$ (called the cross-gradient-Nernst term). The $\gamma$ coefficients are given in \cite{sadler2021}. For in-depth discussion of the impact of Nernst on self-generated magnetic field profiles, see \cite{walsh2017}. For the impact of cross-gradient-Nernst, see \cite{walsh2021a}.
	
	The final terms in equation \ref{eq:mag_trans_new} are the only sources of magnetic flux for ICF implosions. The first, $\nabla \times (\nabla P_e/en_e)$ is the Biermann Battery term, the primary focus of this paper. Often, for demonstration purposes, the ideal gas equation of state is assumed ($P_e = n_e T_e$) and the Biermann Battery term can be re-written as $\nabla T_e \times \nabla n_e/en_e$, i.e. magnetic fields are generated when electron density and temperature gradients are not collinear.
	
	\begin{figure}
		\centering
		\includegraphics[scale=0.45]{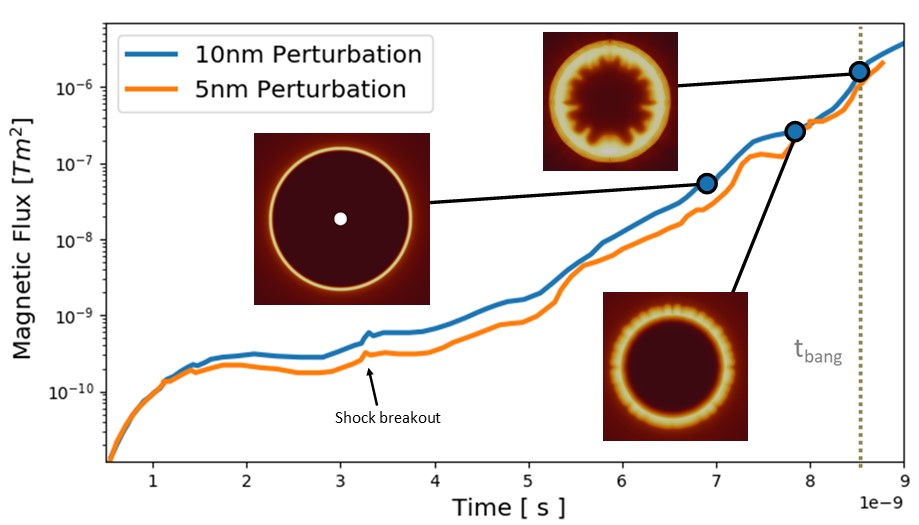}\caption{Magnetic flux $\Phi_B$ as a function of time inside the simulation domain for implosions with two different multi-mode perturbation amplitudes. Approximately 90\% of the magnetic flux is generated in the final 1ns before bang-time, when the asymmetry is greatest and the temperature/density gradients are largest. Density slices are shown at 3 times for the 10nm perturbation case.     \label{fig:flux_vs_time}  }
	\end{figure}
	
	The second source term $-\nabla \times \beta_{\parallel}\nabla T_e/e$ represents magnetic fields generated by gradients in ionization \cite{sadler2020}. This becomes more apparent when it is re-written as $\nabla \beta_{\parallel} \times \nabla T_e / e$; the thermo-electric transport coefficient $\beta_{\parallel}$ only depends on the ionization of the plasma \cite{sadler2021}. This term has been considered when there is mixing of high-Z material into the hot-spot, e.g. by a fill-tube perturbation \cite{sadler2020a}. However, the Biermann Battery term also includes magnetic fields generated by ionization gradients (as $\nabla n_e = \nabla (n_i Z)$); re-writing the total magnetic source (including both Biermann and $\beta_{\parallel}$) using the ideal gas equation gives:
	
	\begin{equation}
		\Bigg[\frac{\partial \underline{B}}{\partial t}\Bigg]_{source} =  \frac{\nabla T_e \times \Big(\frac{\nabla \rho}{\rho} + \nabla Z ( \frac{1}{Z} - \frac{\partial \beta_{\parallel}}{\partial Z} ) \Big)}{e}
	\end{equation}
	
	Here the first two terms are Biermann Battery magnetic fields generated at density gradients and ionization gradients. The $\beta_{\parallel}$ term acts against Biermann generation at ionization gradients, but is always smaller in magnitude. The Biermann fields for a uniform density plasma can only be suppressed by up to 30\% from the $\beta_{\parallel}$ term, which has its greatest contribution at $Z=1$, diminishing to only a 10\% reduction by Z=20. Of course, this is also before the density gradient fields are taken into account in real systems. In this way, when ideal gas is assumed, the $\beta_{\parallel}$ term is only of secondary importance. Nonetheless, magnetic fields generated by the $\beta_{\parallel}$ term are included in all the simulations in this paper. However, the perturbations applied do not create significant mix, making this term's contribution insignificant. Therefore, the $\beta_{\parallel}$ term is not yet included in the presented magnetic flux scaling.

	The simulations in this paper use the Gorgon code \cite{chittenden2004,ciardi2007}, including all of the magnetic transport terms in equations \ref{eq:mag_trans_new} and \ref{eq:mag_trans_new_velocity}. Gorgon has been used previously to study Biermann Battery magnetic field generation, both in ICF capsules \cite{walsh2017} and in laser-driven foils where comparisons to experiments are possible \cite{campbell2020,campbell2021}. The transport coefficients in Gorgon use recently updated dependence on Hall Paramter \cite{sadler2021}, which are an improvement to the Epperlein \& Haines coefficients \cite{epperlein1986}. The new coefficients have been shown to be important in simulations involving self-generated magnetic fields in ICF hot-spots \cite{walsh2021a}. While the discussion in this paper is limited to magnetic field generation and transport, all Gorgon simulations reported here include the impact of magnetic fields on the electron heat-flow, including Righi-Leduc. The thermal transport algorithm is a centered symmetric scheme that is purposely built for anisotropic thermal conduction \cite{sharma2007,walsh2018a}. At no point do the self-generated fields become large enough to directly affect the plasma dynamics through the Lorentz force.
	
	Gorgon's radiation transport uses a $P\frac{1}{3}$ automatic flux-limiting scheme \cite{jennings2006} which captures both the free-streaming and diffusive limits. A thermal and Nernst flux limiters of 0.1 have been chosen, which limit the thermal conduction velocity and Nernst velocity to a tenth of the thermal velocity. No $\alpha-$heating has been included in these simulations. The calculations use a polar mesh during the drive-phase, with \unit[1]{$\mu$m} radial resolution and 360 equally spaced angular zones from $\theta =0,\pi$, where $\theta$ is the polar angle Before the first shock converges onto the axis the simulation is re-mapped to cylindrical geometry with \unit[$\frac{1}{2}$]{$\mu$m} resolution.
	
	The simulations in this paper use the indirect-drive NIF design N170601, which was a cryogenic layered DT implosion from the HDC campaign \cite{clark2019}. As the self-generated magnetic fields require capsule asymmetry to develop, the choice of perturbation is critical. Here HDC shell thickness asymmetries are used \cite{casey2021}. No attempt is made to accurately model the dominant asymmetries in the N170601 experiment; instead, shell thickness asymmetries are applied to the inside surface of the shell to demonstrate the key processes of magnetic flux generation. By varying the applied mode numbers and amplitudes, the scaling of magnetic flux in the hot-spot becomes apparent. First this is done using single-mode perturbations, where the shell thickness variation simply follows a cosine form. More realistic multi-mode simulations then use randomly chosen amplitudes and modes. For example, a $5nm$ multi-mode perturbation  is defined as the maximum peak-to-trough amplitude of each applied mode, with the actual amplitude chosen randomly between 0,5nm. Each mode is chosen randomly between k=1,180 with linear distribution; 400 modes are applied. 
	 
	This paper is organized as follows. Section \ref{sec:flux_vs_time} characterizes the magnetic flux in the capsule as a function of time. For the perturbation sources simulated here, the magnetic flux in the hot-spot is dominated by generation during hot-spot formation. Section \ref{sec:flux_stag_single} uses single-mode simulations to as a starting point for the magnetic flux scaling. Section \ref{sec:flux_stag_multi} then extends this to multi-modal situations, and allows for estimates of flux as a function of $\theta$ in the capsule. Section \ref{sec:transp_stag} then broadens the discussion to transport of magnetic flux within the hot-spot. The magnetic flux model developed in section \ref{sec:flux_stag_multi} is extended to assume Nernst deposits the flux at the hot-spot edge over a finite width, thereby completing a model that takes radiation hydrodynamics data as an input and reproduces extended-MHD simulation results. Finally, the appendix shows the limitations of the model for large perturbation sources and P1 drive asymmetries that result in target offsets.
	 
	 In this paper 'simulated flux' refers to the magnetic flux calculated using 2D extended-MHD simulations, while 'model flux' refers to the expectation given by the scaling.
	 
	\section{Magnetic Flux Generation Throughout Implosion \label{sec:flux_vs_time}}

	This section is a general overview of magnetic flux generation across the whole capsule implosion. The standard definition of magnetic flux is:
	
	\begin{equation}
	\Phi = \iint_S  \underline{B} \cdot  \delta \underline{S} \label{eq:phi_standard}
	\end{equation}
	
	Where $S$ is a chosen surface, e.g. a 2-D slice through a hot-spot.	Equation \ref{eq:phi_standard} is not an adequate definition for the purposes of this paper. The inadequacy is made clear by looking at how $\Phi$ grows due to the Biermann Battery term and applying Stokes' theorem:
	
	\begin{equation}
	\Bigg[\frac{\partial \Phi}{\partial t} \Bigg]_{Biermann}= \oint_{\delta S}  \frac{\nabla P_e}{en_e} \cdot \delta \underline{l} \label{eq:dphi_dt}
	\end{equation}
	
	Taking a slice through a capsule, the surface is defined (in spherical co-ordinates) by the range $r=[0,\inf],\theta=[0,2\pi]$. The total magnetic flux is always zero ($\nabla P_e = 0$ at large radius). 
	
	Another surface of interest is just the 2D surface simulated, in the range $r=[0,\inf],\theta=[0,\pi]$. Here, the line integral in equation \ref{eq:dphi_dt} is only dependent on the pressure gradient and electron density along the axis of symmetry. Therefore, for this surface the total flux is only affected by a mode 1 asymmetry. This makes intuitive sense, as a mode 2 asymmetry will generate regions of both positive and negative $B_\phi$, which cancel out under the definition given in equation \ref{eq:phi_standard}. This paper is not concerned with the overall cancellation of flux vectors; instead, quantification of how much of the plasma has an embedded magnetic field is sought. 
	
	Throughout this paper, the following alternative definition is used for magnetic flux:
	\begin{equation}
	\Phi_B = \iint_S | \underline{B} \cdot \underline{\hat{n}} |  \delta S \label{eq:flux}
	\end{equation}
	where $\underline{\hat{n}}$ is the surface normal. Calculating $\Phi_B$ is simple in a 2D simulation, as the surface of interest is the whole simulation domain. In addition, 2D simulations only generate a $B_{\theta}$ magnetic field component. In 3D, however, all components of magnetic field can be generated, and a choice must be made of which surface to calculate the flux over. 
	
	Self-generation of magnetic fields by the Biermann Battery mechanism is intrinsically related to the symmetry of the capsule implosion. Without perturbations, no magnetic fields are generated. Larger temperature and logarithmic electron densities also generate more magnetic field. 
	
	Figure \ref{fig:flux_vs_time} shows the magnetic flux ($\Phi_B$) evolution as a function of time for multi-mode simulations with $5nm$ and $10nm$ initial amplitude perturbations. The distribution of modes is identical between the 5nm and 10nm cases, with just the shell thickness amplitudes modified. For reference, the first shock breaks out of the HDC shell at around 3.3ns, and is indicated in the figure. The first shock then converges on axis by 7.75ns, beginning the stagnation phase. The time of peak neutron production is then 8.5ns.
	
	Figure \ref{fig:flux_vs_time} also includes density profiles at 3 times to demonstrate the perturbation evolution. At the earlier time, 6.9ns, the effect of the thickness variations cannot be detected with the naked eye. At this point the first shock has not yet converged onto the target chamber center. While the temperature of the compressed gas on the inside of the shell can reach hundreds of eV, the relative symmetry of the implosion at this time does not generate much magnetic flux. Approximately 90\% of the magnetic flux is generated in the final 1ns of the implosion, when the temperature and density gradients are largest, and the asymmetries have grown non-linear.
	
	\begin{figure}
		\centering
		\begin{subfigure}[b]{0.5\textwidth}
			\centering
			\includegraphics[width=1.\textwidth]{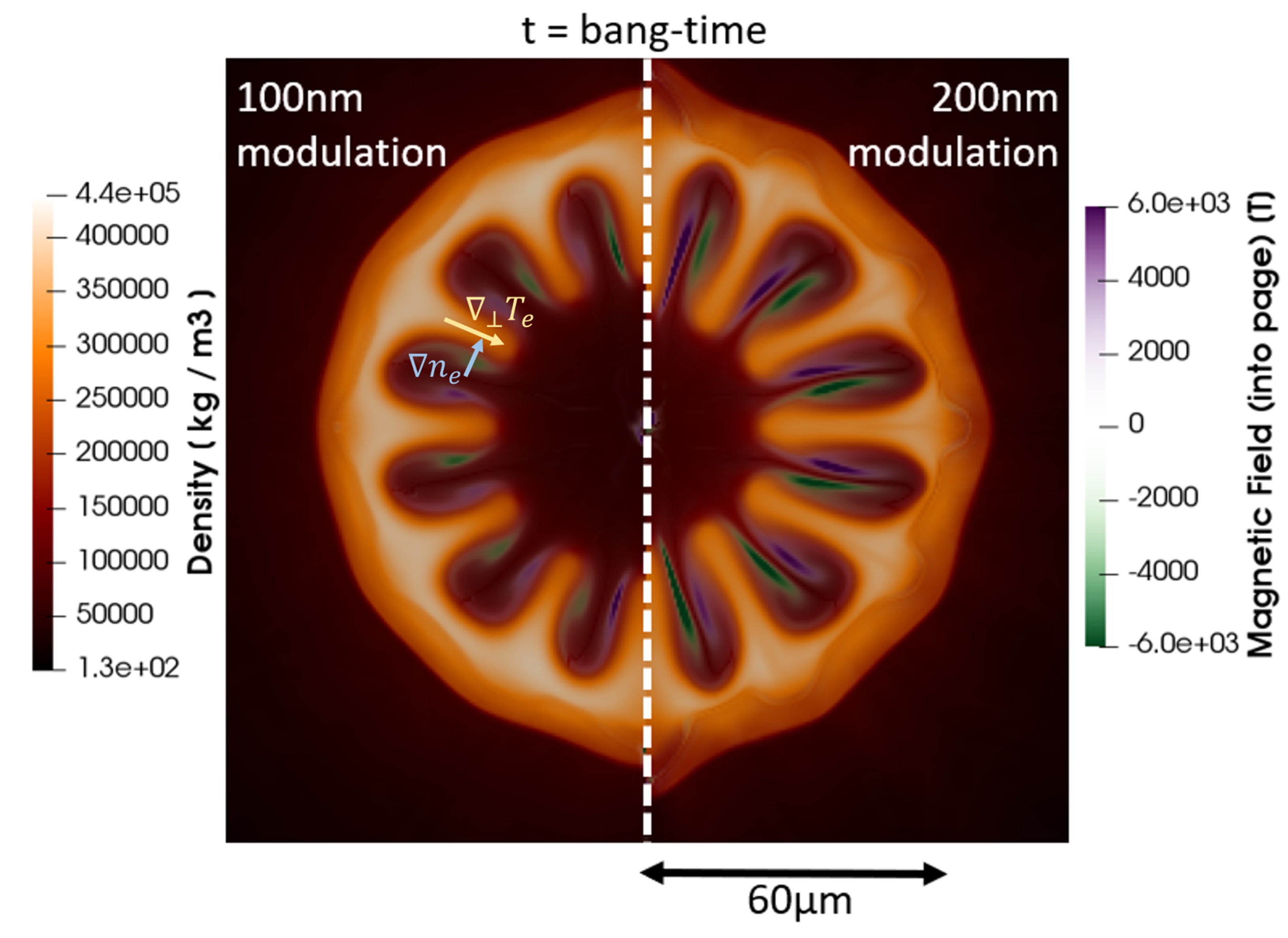}
			\caption{Density and (overlaid) magnetic field profiles at bang-time for capsules with mode 12 perturbations. The simulation on the left was initialized with $100nm$ shell thickness variations, while the simulation on the right used $200nm$. }
			\label{fig:hs_field_amp}
		\end{subfigure}
		\begin{subfigure}[b]{0.5\textwidth}
			\centering
			\includegraphics[width=1.\textwidth]{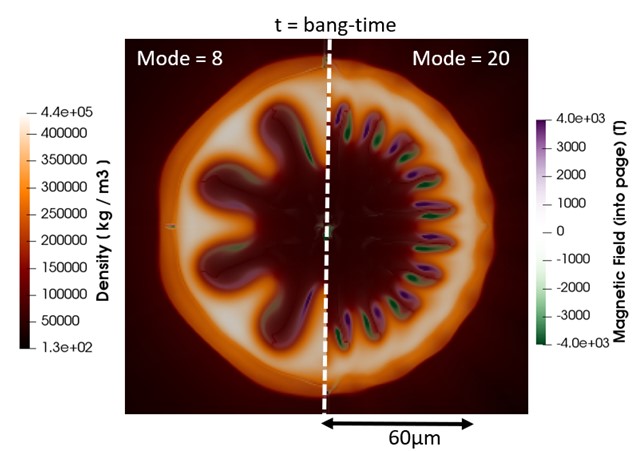}
			\caption{Density and (overlaid) magnetic field profiles at bang-time for capsules with $100nm$ initial shell thickness variations. The simulation on the left is for a mode 8 perturbation, and the simulation on the right uses a mode 20.}
			\label{fig:hs_field_mode}
		\end{subfigure}
		
		\caption{}
		\label{fig:hs_field}
	\end{figure}
	
	Concerns have been raised about using Biermann Battery algorithms at shock fronts \cite{graziani2016}, due to the so-called 'Biermann catastrophe', where the computed magnetic field strength does not converge with radial cell size. In this work the shell is being perturbed, so the shock front remains predominantly symmetric and no numerical issues arise. Future work will study in detail the flux generated at perturbed shock-fronts, demonstrating that no numerical issues arise unless the self-generated fields are large enough to magnetize the thermal front.
	
	For indirect-drive the temperature gradients in the ablation region are small, resulting in weak magnetic fields and insignificant electron magnetizations. Even if the heat-flow were magnetized, the transport of energy is dominated by radiation, which is not suppressed by magnetic fields. For direct-drive, however, even small magnetizations can result in changes to perturbation growth \cite{hill2017}, despite the Biermann term being suppressed below its classical value \cite{sherlock2020}. Long time-scale extended-MHD simulations of this process are underway, with a brief overview given within \cite{sadler2021}.
	
	%\note{No need to talk about Biermann catastrophe. Cite Graziani and then discuss that there isn't much perturbation of the shock front here. Behind shock front we have field strengths of less than 1T. Future publication with perturbed shock front will look at this more closely.}
	
	%Appendix goes into more detail, but conclusion is that total magnetic flux generation is converged with cell size. But field strength generated each timestep is not converged. Therefore, as long as the field generation does not affect the shock front too much we have a converged simulation. 
	
	\section{Magnetic Flux Generated during Stagnation: Single-Mode Perturbations \label{sec:flux_stag_single}}
	
	%\note{another way of looking at it: For a given amplitude, around each spike the same amount of flux is generated. But you fit more high mode spikes into a whole capsule}

	The previous section demonstrated that magnetic flux generation is typically greatest during the hot-spot stagnation. This section outlines how the magnetic flux generation in the hot-spot scales with bulk hot-spot quantities as well as perturbation amplitudes and modes, focusing on capsules perturbed by a single mode. The subsequent section extends the theory here to more realistic multi-mode implosions.
	
	First, a scaling is demonstrated qualitatively. Figure \ref{fig:hs_field_amp} shows two mode 12 density profiles at bang-time, one with an initial 100nm shell thickness variation and another with a 200nm variation. The magnetic field strength is plotted over the density. The direction of electron density gradient is shown with a light blue arrow, as well as the component of the temperature gradient that is perpendicular to the blue arrow. As a spike pushes into the hot-spot, the tip is heated up faster than the base, which result in the field topology shown. 
	
	The total magnetic flux increases from $4.3\times 10^{-6}$Tm$^2$ to $8.0\times 10^{-6}$Tm$^2$ by increasing the amplitude; this makes intuitive sense, as an implosion with no applied perturbation results in no hot-spot magnetic flux, as $\nabla T_e \times \nabla \log n_e = 0$. While this general trend is true here, it is worth noting that if the perturbations are large enough to substantially lower the hot-spot temperature, this can also lower the Biermann generation. 
	
	Figure \ref{fig:hs_field_mode} compares the density and magnetic field strength of a mode 8 to a mode 20 simulation. Both perturbations were initialized with a 100nm amplitude. By bang-time the mode 20 perturbation is smaller than the mode 8 due to enhanced thermal ablative stabilization \cite{betti2001}. Nonetheless, the field generation of the higher mode is greater, with a total magnetic flux of $4.2\times 10^{-6}$Tm$^2$ compared with $2.7\times 10^{-6}$Tm$^2$. This can be explained by the non-radial temperature and density scale lengths being shorter.
	
	So, for a given perturbation size, higher modes give more magnetic flux. But, for a given perturbation mode number, larger perturbations give more magnetic flux.	A general scaling for this begins by decomposing the Biermann Battery generation term:

	\begin{equation}
		\Bigg[\frac{\partial B_{\phi}}{\partial t}\Bigg]_{Biermann} = \frac{1}{e r} \Bigg( \frac{\partial T_e }{\partial r} \frac{\partial \ln n_e }{\partial \theta} - \frac{\partial \ln n_e }{\partial r} \frac{\partial T_e }{\partial \theta}\Bigg) \label{eq:dbier}
	\end{equation}

	Where the hot-spot plasma is assumed to obey the ideal gas law, $P_e = n_e T_e$. The temperature and density length-scales in $\theta$ are shorter for higher mode perturbations:
	
	\begin{equation}
		\Bigg|\frac{\partial T_e}{\partial \theta} \Bigg| = \frac{k \Delta_{\theta} T_e}{2 \pi} \label{eq:dT_dtheta}
	\end{equation}
	\begin{equation}
	\Bigg|\frac{\partial  \ln n_e}{\partial \theta} \Bigg| = \frac{k \Delta_{\theta} \ln n_e}{2 \pi}  \label{eq:dn_dtheta}
	\end{equation}

	\begin{figure}
		\centering
		\begin{subfigure}[b]{0.5\textwidth}
			\centering
			\includegraphics[width=1.\textwidth]{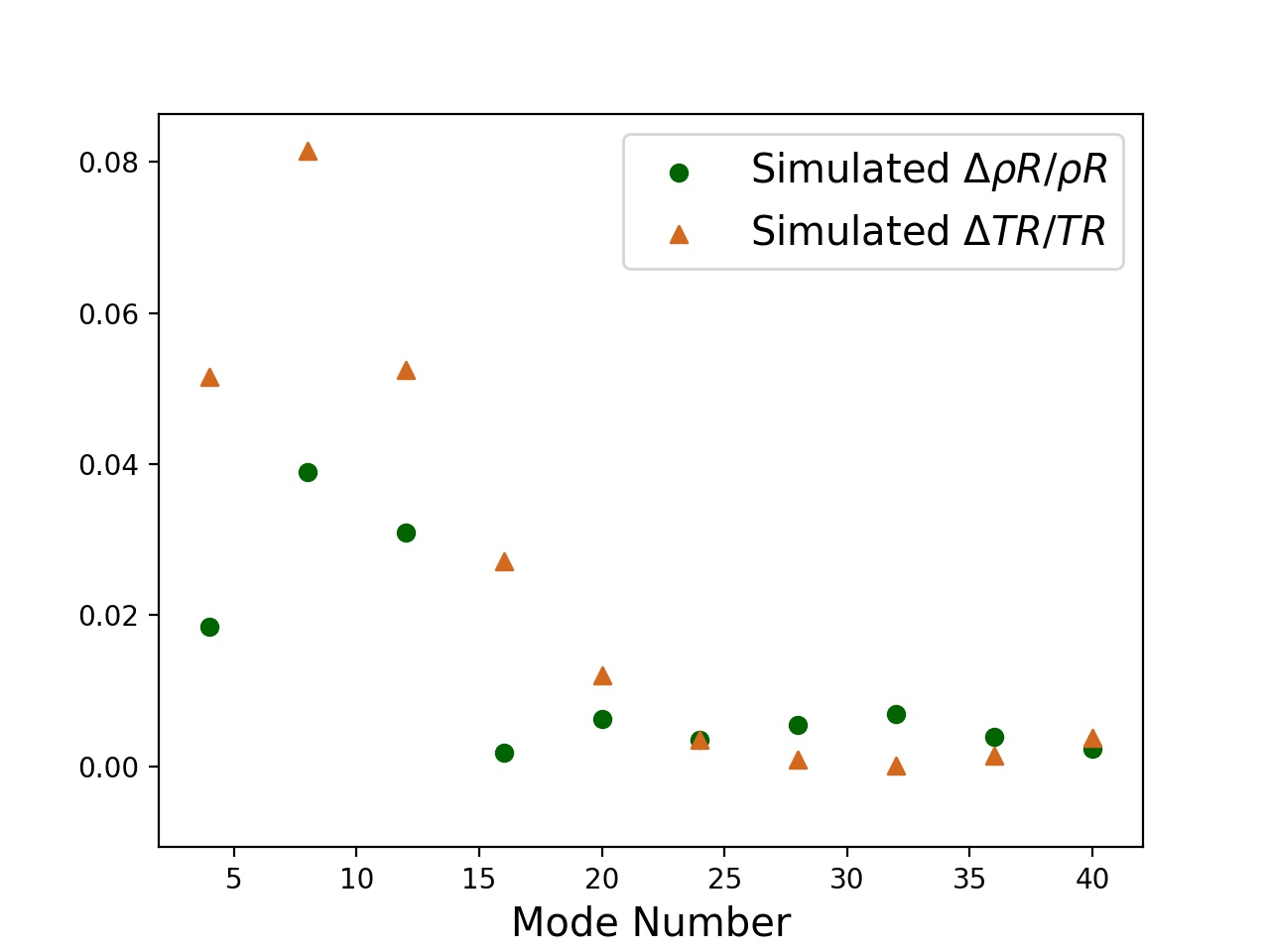}
			\caption{Normalized $\rho R$ (radially-integrated density) and $TR$ (radially integrated temperature) perturbations at bang-time for simulations with single-modes applied. Time resolved versions of this data is used to predict the magnetic flux as a function of mode number below. }
			\label{fig:singlemode_rhoR}
		\end{subfigure}
		\begin{subfigure}[b]{0.5\textwidth}
			\centering
			\includegraphics[width=1.\textwidth]{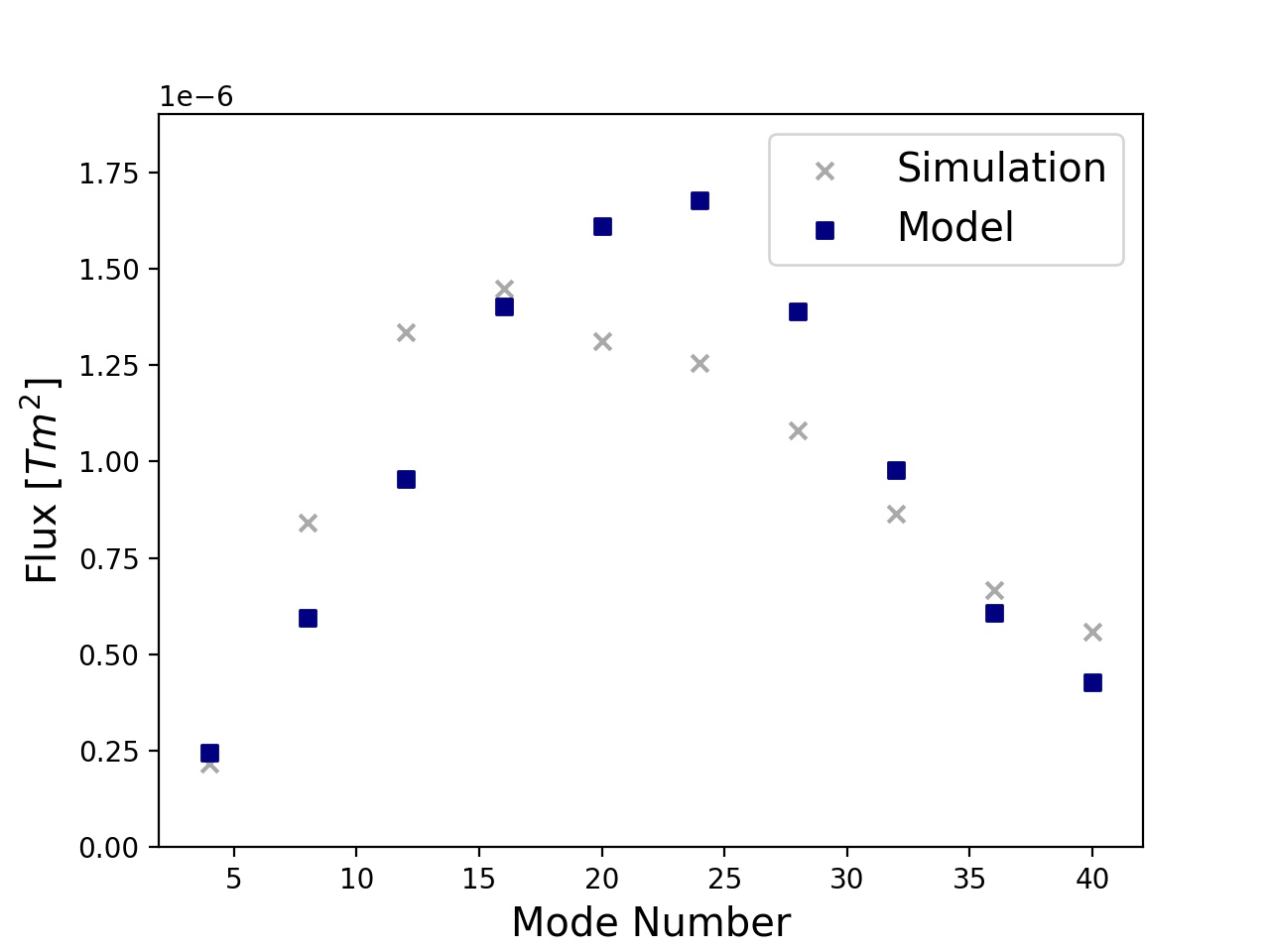}
			\caption{Magnetic flux as a function of mode number for single-mode capsule simulations initialized with 100nm shell thickness variations. The model predictions here are based on equation \ref{eq:scaling_singlemode}.}
			\label{fig:singlemode_flux}
		\end{subfigure}
		
		\caption{Model input (a) and comparison of model output to extended-MHD simulations (b). Each point represents a different single-mode simulation. Even though high-modes have been ablatively stabilized by bang-time (a) there can still exist large magnetic fluxes on these scales (b).}
		\label{fig:singlemode}
	\end{figure}
	
	Where $k$ is the perturbation mode number. Both the perturbation amplitude and mode number dependence can be seen in equations \ref{eq:dT_dtheta} and \ref{eq:dn_dtheta}. For a larger perturbation, $\Delta_{\theta} T_e$ and $\Delta_{\theta} \ln n_e$ 	are larger. It is important to note that $\Delta_{\theta} T_e$ and $\Delta_{\theta} \ln n_e$ are both functions of $k$, which can clearly be seen in the density of figure \ref{fig:hs_field_mode}, with the mode 20 implosion much less perturbed than the mode 8.
	
	The scaling of Biermann generation rate (equation \ref{eq:dbier}) is simplified by looking at the total magnetic flux (from equation \ref{eq:flux}) for a single-mode perturbation:
	
	\begin{equation}
		\Bigg[\frac{\partial \Phi_B}{\partial t} \Bigg]_{singlemode}= \frac{k}{e} \Big( (T_{hot} - T_{shell}) \Delta_{\theta} \ln n_e + \ln \frac{n_{shell}}{n_{hot}}\Delta_{\theta} T_e\Big)
	\end{equation}
	
	Where $T_{hot}$ and $n_{hot}$ are the electron temperature and density in the hot-spot center and $T_{shell}$ and $n_{shell}$ are the electron temperature and density in the shell.
	
	To advance, a quantification of the perturbation in $\theta$ is required. For this, the line-integrated density ($\rho R$) and temperature ($TR$) are used. An ansatz is made that the density and temperature jumps take the form $\Delta_{\theta} T_e = \frac{\Delta TR}{TR} (T_{hot}-T_{shell})$ and $\Delta_{\theta} n_e = \frac{\Delta \rho R}{\rho R} (n_{shell}-n_{hot})$; this will be verified in the comparisons below. $\Delta TR$ and $\Delta \rho R$ are the variations in line-integrated density and temperature due to the mode number $k$. The magnetic flux generation due to Biermann becomes:
	
	\begin{equation}
	\begin{split}
		\Bigg[\frac{\partial \Phi_B}{\partial t} \Bigg]_{singlemode}= \\
		k \ln \Big( \frac{n_{shell}}{n_{hot}} \Big) \frac{T_{hot} - T_{shell}}{e} \Bigg(\frac{\Delta \rho R}{\rho R} - \frac{\Delta T R}{T R} \Bigg) \label{eq:scaling_singlemode}
	\end{split}
	\end{equation}
	
	This scaling can be tested against extended-MHD simulations by inputting rad-hydro data ($n_{hot}, n_{shell}, T_{hot}, T_{shell}, \Delta \rho R, \rho R, \Delta T R, TR$) as a function of time into the model and comparing the predicted magnetic flux to that calculated within the MHD routines of the simulation. 
	
	Hot-spot bulk quantities ($n_{hot}$ and $T_{hot}$) are taken as the time-resolved burn-averaged electron density and temperature. The shell quantities ($n_{shell}$ and $T_{shell}$) are taken as the mass-averaged electron density and temperature within the fuel at a given time. Radial integrals of the density and temperature ($\rho R$ and $TR$) are taken from the hot-spot center to $n_e = n_{shell}$ and $T_e = T_{shell}$ respectively. $\Delta \rho R(t)$ and $\Delta T R(t)$ values are calculated by taking the Fourier transform of $\rho R(\theta,t)$ and $T R(\theta,t)$ and taking the applied mode component; in this way the perturbation is averaged over the whole simulation, rather than a single spike. Issues with tracking the hot-spot center when a large mode 1 perturbation is applied are discussed in appendix \ref{sec:model_tests}. The model is initialized at $t=8.1$ns, when the first shock has already rebounded off the axis and has reached the shell, signifying the beginning of the stagnation phase. The time-step used in the model is 50ps, which is more than a factor of 100 greater than the MHD simulation time-step.
	
	Figure \ref{fig:singlemode_rhoR} shows  simulated $\frac{\Delta \rho R}{\rho R}$ and $\frac{\Delta T R}{T R}$ for single mode calculations at bang-time. The higher modes have almost completely been suppressed at this time, with $k \ge 20$ exhibiting $\frac{\Delta \rho R}{\rho R} \approx 0$.
	
	Figure \ref{fig:singlemode_flux} then shows the magnetic flux in the capsule at bang-time. Both the extended-MHD simulation results and the model predictions are shown for comparison. For low mode numbers $k \le 16$ the model under-predicts the magnetic flux. For higher modes, the full 2-D simulations expect less flux than the model.
	
	To be clear, the time-dependent $\rho R$ and $TR$ variations are taken from the simulations and are used as inputs to the model, along with bulk hot-spot and shell temperatures/densities. The model itself does not have the ability to predict hydrodynamic growth rates; it uses the simulated radiation-hydrodynamic perturbation sizes to predict magnetic flux generation. A favorable comparison of magnetic flux between the model and simulations does not validate the perturbation growths calculated.
	
	Figures \ref{fig:singlemode_rhoR} and \ref{fig:singlemode_flux} demonstrate an important aspect of magnetic flux generation: even if high mode perturbations have been suppressed by bang-time, it does not mean that the magnetic fields associated with those perturbations do not still exist. The hot-spot contains the time history of all perturbations that have developed. While $\Delta \rho R /\rho R < 1\%$ for $k=20$ at bang-time, the magnetic flux in the hot-spot is still comparable to the flux generated by the $k=12$ simulation, which still has a highly perturbed hot-spot at bang-time.

	\section{Magnetic Flux Generated During Stagnation: Multi-Mode Perturbations \label{sec:flux_stag_multi}}
	
	\begin{figure}
		\begin{subfigure}[b]{0.5\textwidth}
		\centering

		\includegraphics[scale=0.5]{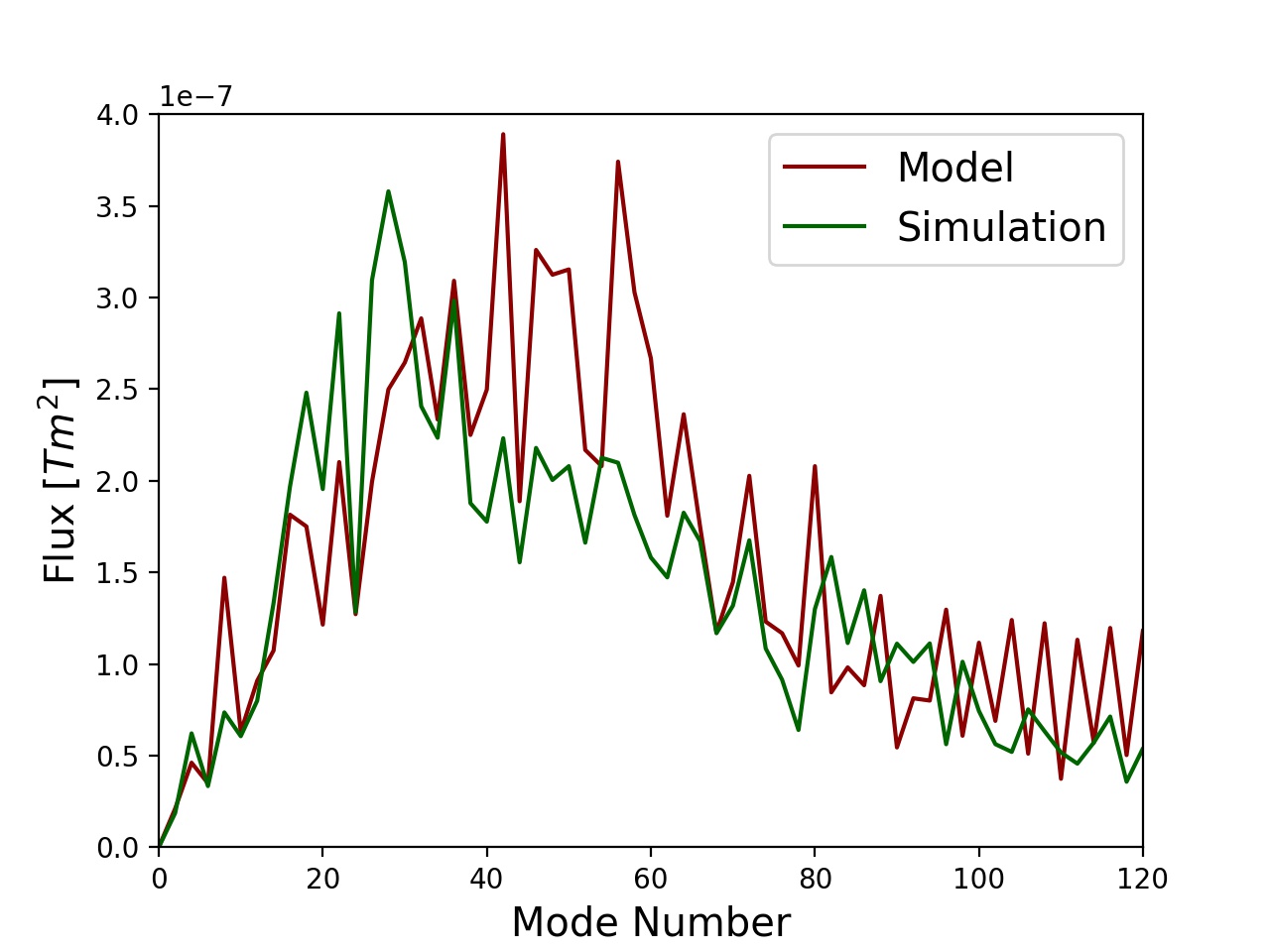}\caption{ \label{fig:flux_vs_mode}  Averaged magnetic flux distribution for the theoretical model compared with the simulations. }
		\end{subfigure}
		\begin{subfigure}[b]{0.5\textwidth}
			\centering
			\includegraphics[width=1.\textwidth]{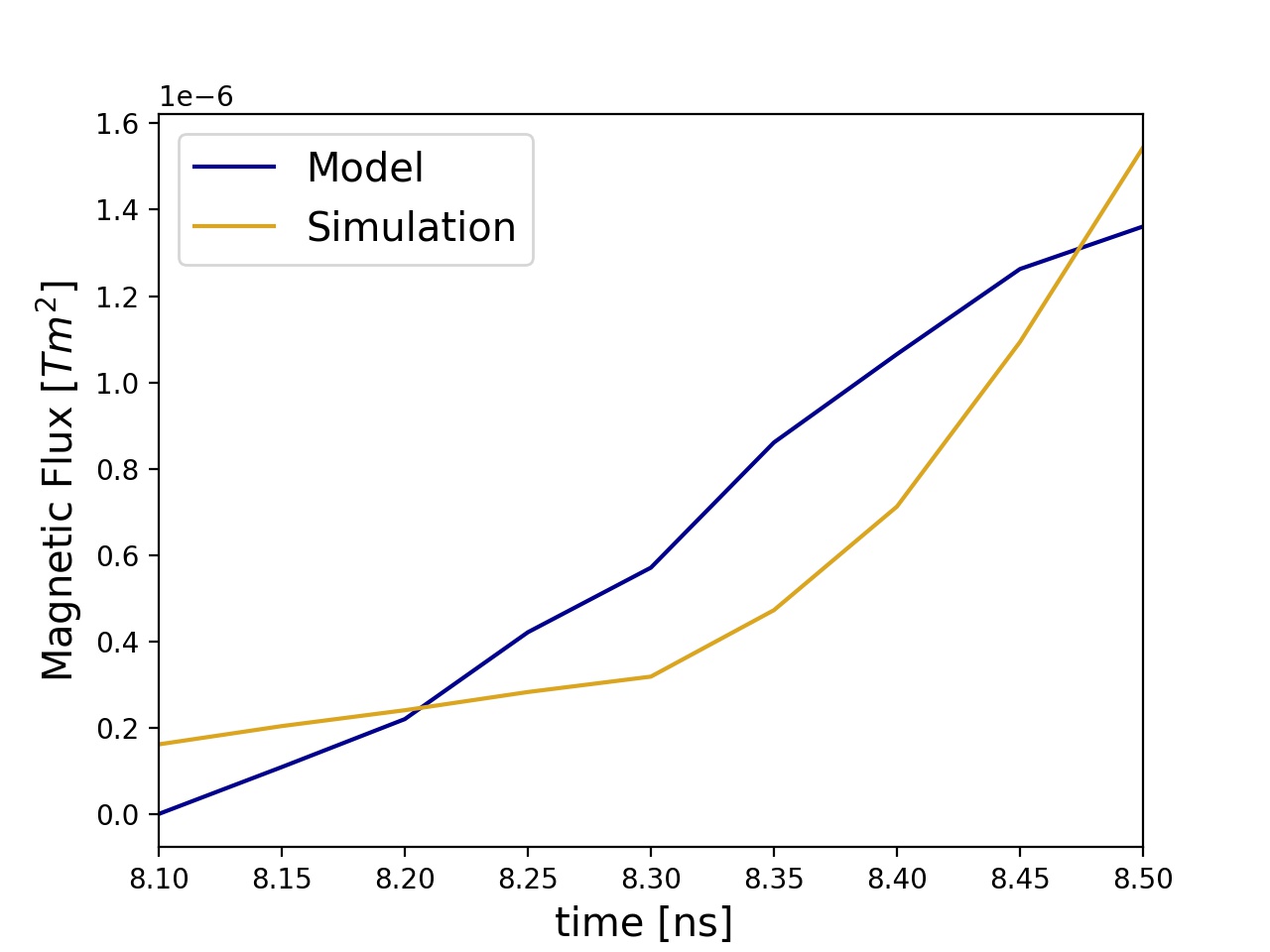}
			\caption{Temporal evolution of magnetic flux for the theoretical model and the simulations. The model over-predicts early-time magnetic flux generation rates, which explains why the high-mode features (modes 35-60) are over-estimated in figure \ref{fig:flux_vs_mode}.}
			\label{fig:Flux_vs_time_model}
		\end{subfigure}
	\caption{Comparison of theoretical model to full simulations for 8 averaged multi-mode simulations initialized with different random numbers used to generate the perturbations. The model uses the density and temperature evolution from the simulations to predict the flux distribution.}
	\end{figure}
	
	Equation \ref{eq:scaling_singlemode} can be extended to capsules perturbed with a range of modes by decomposing the magnetic flux into individual mode number contributions:
	
	\begin{align}
	\begin{split}
		\Phi_B(k,t)= i \int_{t_0}^{t} k \ln \Big( \frac{n_{shell}(t)}{n_{hot}(t)} \Big) \frac{T_{hot}(t) - T_{shell}(t)}{e} \\
		\Bigg(\frac{\rho R(k=k,t)}{\rho R(k=0,t)} - \frac{T R(k=k,t)}{T R(k=0,t)} \Bigg) \delta t \label{eq:scaling}
	\end{split}
	\end{align}
	
	Where the phase information of the magnetic flux is retained. The imaginary number $i$ represents a shift in phase of $\pi /2$ from the $\rho R$ and $TR$ perturbations. This is made clear by looking at the magnetic field distributions in figures \ref{fig:hs_field_amp} and \ref{fig:hs_field_mode}; the peak field strength is out of phase with the peak density (or $\rho R$) locations. 
	
	To test this scaling, the $10nm$ multi-mode perturbation shown in figure \ref{fig:flux_vs_time} was run 8 times using different random number seeds. Averaging the results smooths out the flux as a function of mode number. As in section \ref{sec:flux_stag_single}, the model is run on the data once the deceleration phase begins; this is defined as the time when the first shock has bounced off the axis and returns to the shell.
	
	Figure \ref{fig:flux_vs_mode} compares the model prediction for magnetic flux as a function of mode number to the simulated flux distribution for the 8 averaged simulations. There is broad agreement, although the model over-predicts the flux generated for mode numbers in the range $k=[35,60]$.
	
	Figure \ref{fig:Flux_vs_time_model} then shows the temporal evolution of total magnetic flux for the simulation ensemble in comparison with the model. The model is initialized at 8.1ns, so there is no predicted flux before this time. The model clearly over-predicts the magnetic flux generation rate early in the hot-spot formation and under-predicts in the late phase. This is likely associated with the definitions of bulk hot-spot quantities utilized. The enhanced generation early in time explains the excess of magnetic flux predicted for modes $k=[35,60]$ in figure \ref{fig:flux_vs_mode}; earlier in time the higher modes are more unstable than at later times. Likewise, the lower predicted generation rate later in times explains why the model under-predicts the lower mode flux contributions in figure \ref{fig:flux_vs_mode}; these modes continue to have large amplitudes around bang-time.
	
\begin{figure}
		\centering
		\begin{subfigure}[b]{0.5\textwidth}
			\centering
			\includegraphics[width=1.\textwidth]{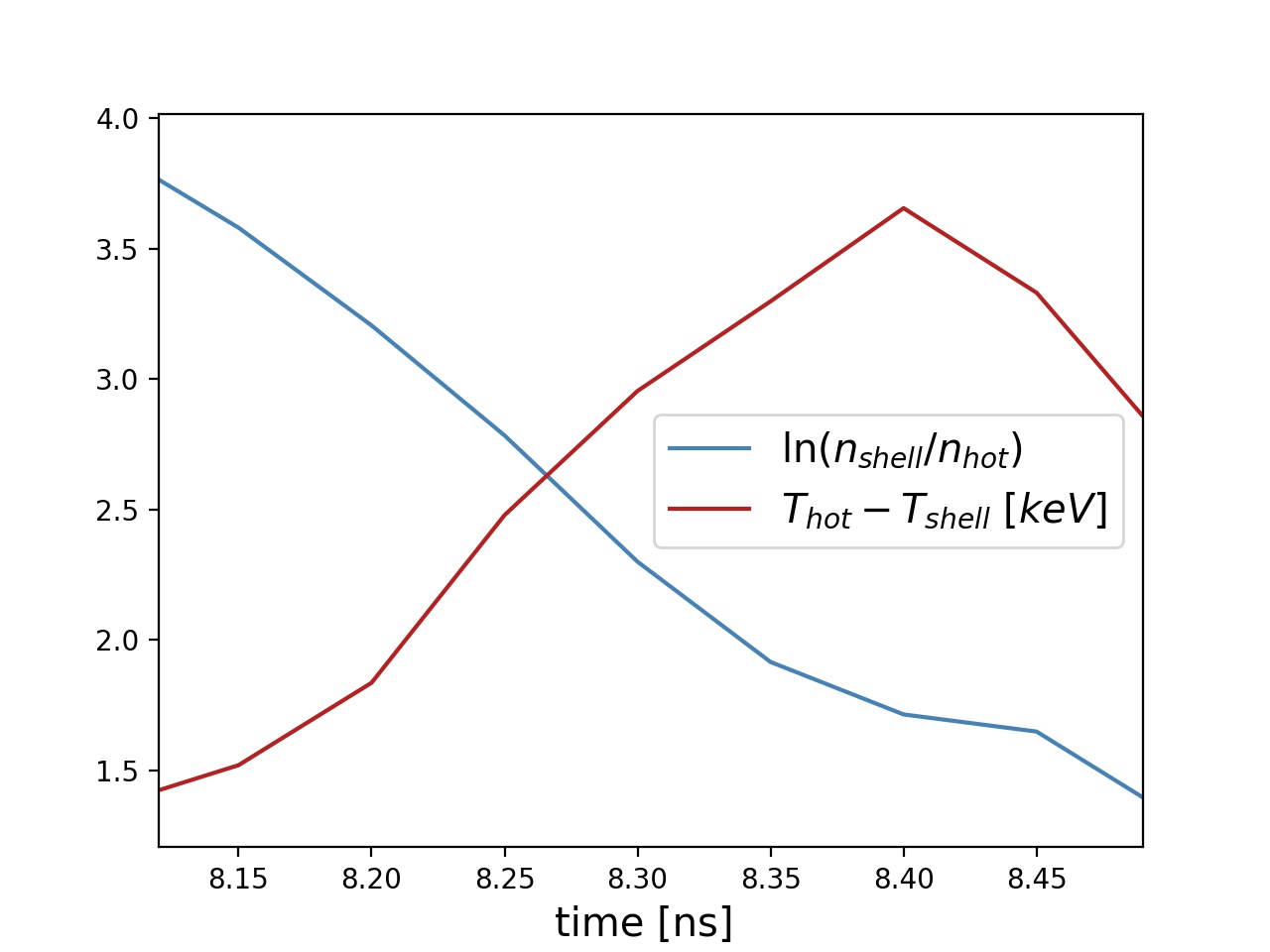}
			\caption{Temporal dependence of the temperature and logarithmic density differentials between the hot-spot and the shell. These represent the overall gradients driving the magnetic flux generation.}
			\label{fig:Temporal_bulk}
		\end{subfigure}
		\begin{subfigure}[b]{0.5\textwidth}
			\centering
			\includegraphics[width=1.\textwidth]{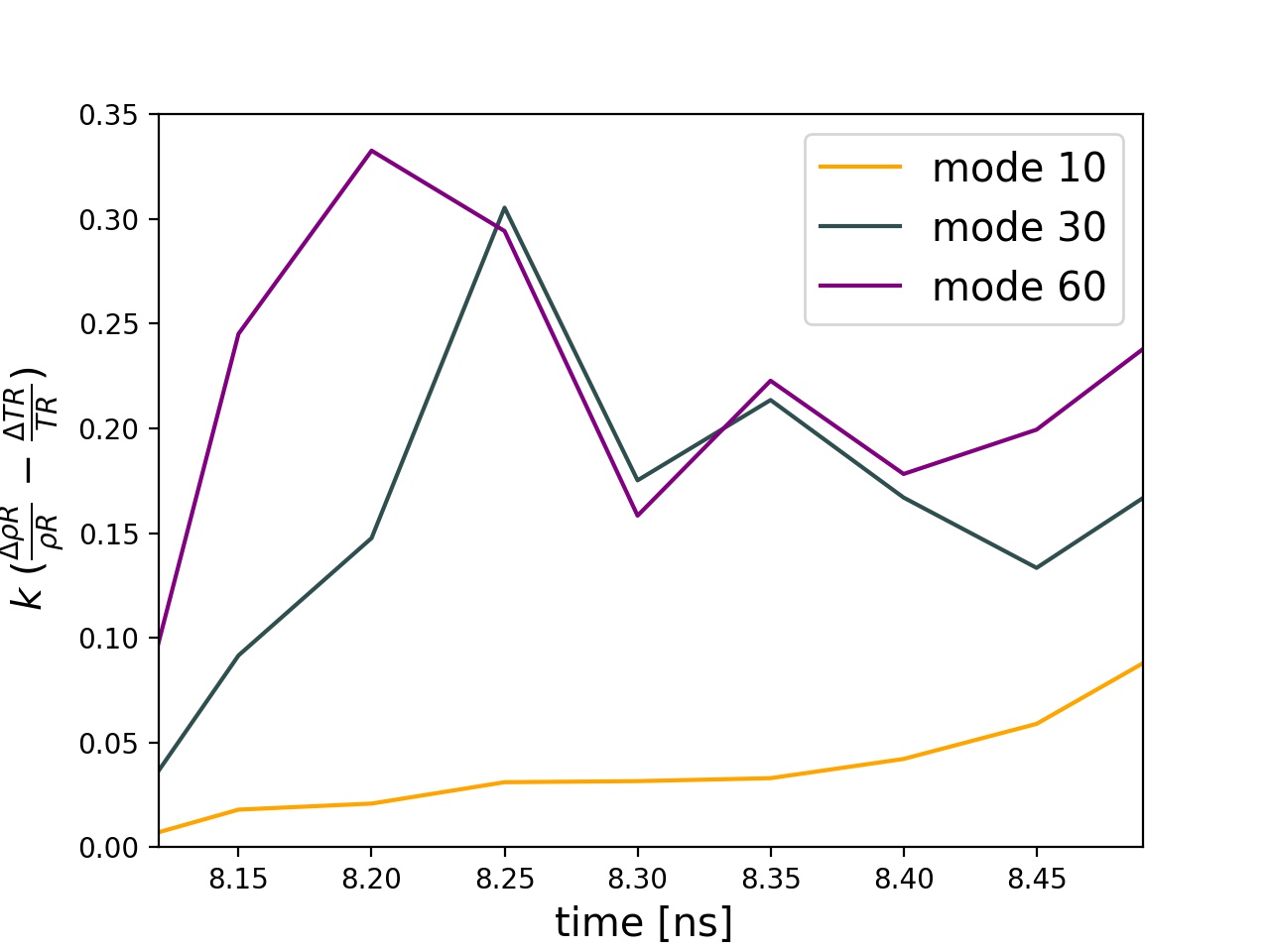}
			\caption{$k (\frac{\rho R(k=k)}{\rho R(k=0)} - \frac{TR(k=k)}{TR(k=0)}) $ for $k=10,30,60$. This factor represents the size of the perturbation driving magnetic flux generation.}
			\label{fig:Temporal_modes}
		\end{subfigure}
		\caption{Inputs into the model (equation \ref{eq:scaling} used to generate the model prediction in figure \ref{fig:flux_vs_mode}.}
		\label{fig:Temporal}
	\end{figure}

	The model can be used to further understand what drives the temporal evolution of magnetic flux in an ICF hot-spot. Figure \ref{fig:Temporal} displays the different terms in equation \ref{eq:scaling} as a function of time.
	
	Figure \ref{fig:Temporal_bulk} shows the bulk temperature and density differentials that affect magnetic field generation. These metrics drive magnetic flux growth equally across all mode numbers. At early times the hot-spot is cool, but the difference in density is large (around a factor of 40). At later times the hot-spot temperature is large (with temperature differences of up to 3.5keV) but the core hot-spot is a more similar density to the shell. These two effects approximately cancel out one another, and the product of $\ln(n_{shell}/n_{hot})(T_{hot}-T_{shell})$ is approximately constant for the hot-spot deceleration time. Note that this result is highly dependent on implosion design, and quantification of this product could be used to inform on the importance of self-generated magnetic fields for a given implosion. 
	
	Figure \ref{fig:Temporal_modes} shows $k (\rho R(k=k)/\rho R(k=0) - T R(k=k)/T R(k=0))$ as a function of time for mode numbers 10, 30 and 60. While the hot-spot quantities $\ln(n_{shell}/n_{hot})$ and $T_{hot} - T_{shell}$ plotted in figure \ref{fig:Temporal_bulk} represent the bulk gradients driving the Biermann Battery generation, the term $k (\rho R(k=k)/\rho R(k=0) - T R(k=k)/T R(k=0))$ represents how non-colinear are the temperature and density gradients. 
	
 	Figure \ref{fig:Temporal_modes} indicates that low mode perturbations ($k=10$ in this case) develop throughout the hot-spot stagnation, with peak asymmetry in the temperature and density gradients at the time of peak neutron production ($t=8.5$ns). For the higher modes, time of peak flux production moves earlier, as the perturbations are stabilized by thermal ablative stabilization.
	
\begin{figure}
		\centering

			\centering
		\includegraphics[scale=0.5]{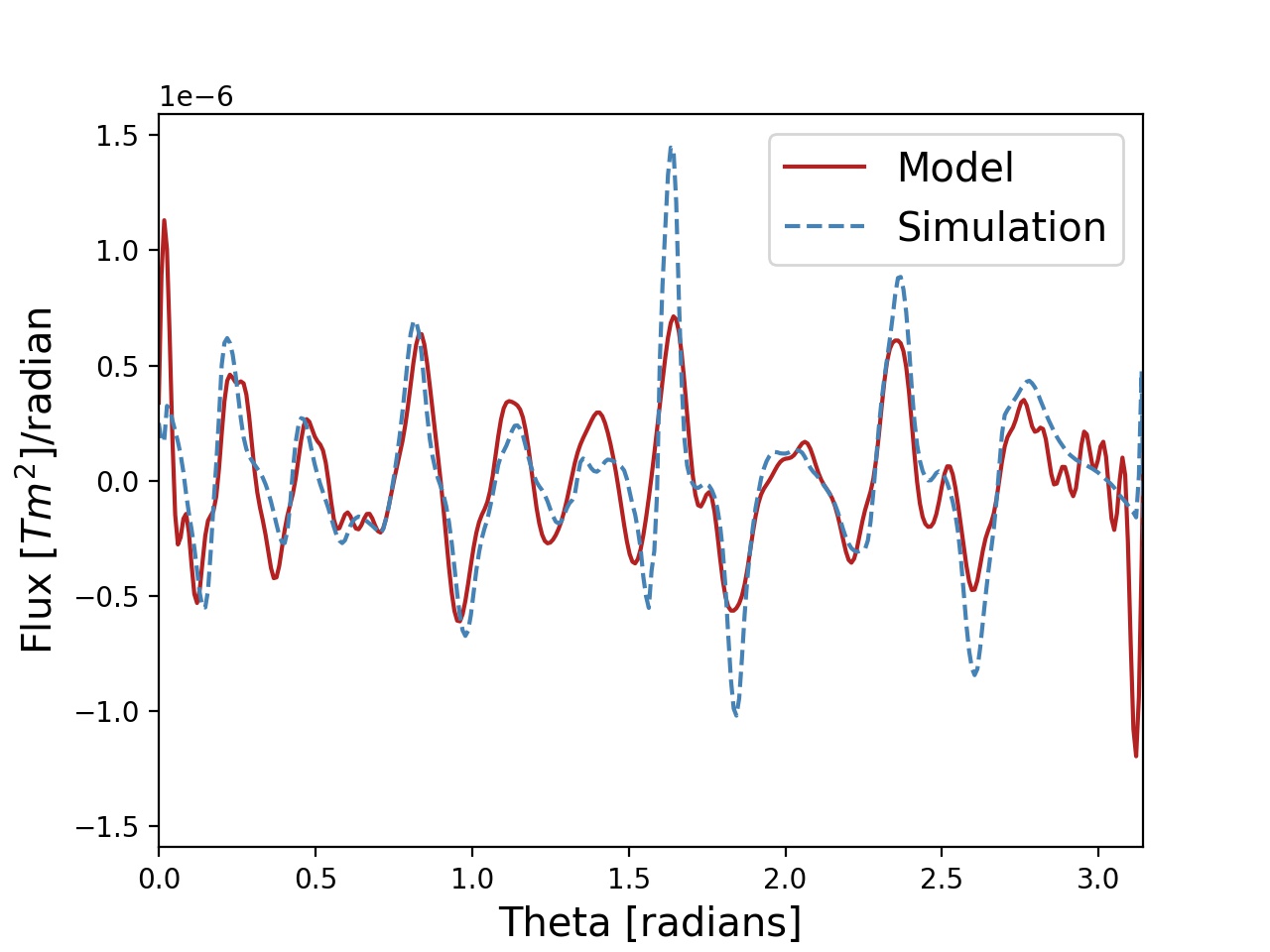}\caption{ \label{fig:flux_vs_theta} Predicted and simulated magnetic flux as a function of $\theta$ for a multi-mode simulation. }

	\end{figure}

	By retaining the phase information in equation \ref{eq:scaling}, it is possible for the model to estimate the magnetic flux as a function of $\theta$. Figure \ref{fig:flux_vs_theta} demonstrates this for one of the $10nm$ multi-mode perturbation simulations. Considering the only spatially-resolved inputs to the model are $\rho R(\theta,t)$ and $TR(\theta,t)$, the model closely captures the location of the flux. The model assumes that there is no non-radial motion of flux; clearly for highly perturbed capsules this assumption breaks down. This is investigated in appendix \ref{sec:model_tests}

	\section{Magnetic Transport during Stagnation \label{sec:transp_stag}}

	\begin{figure}
	\centering
	\begin{subfigure}[b]{0.5\textwidth}
		\centering
		\includegraphics[width=1.\textwidth]{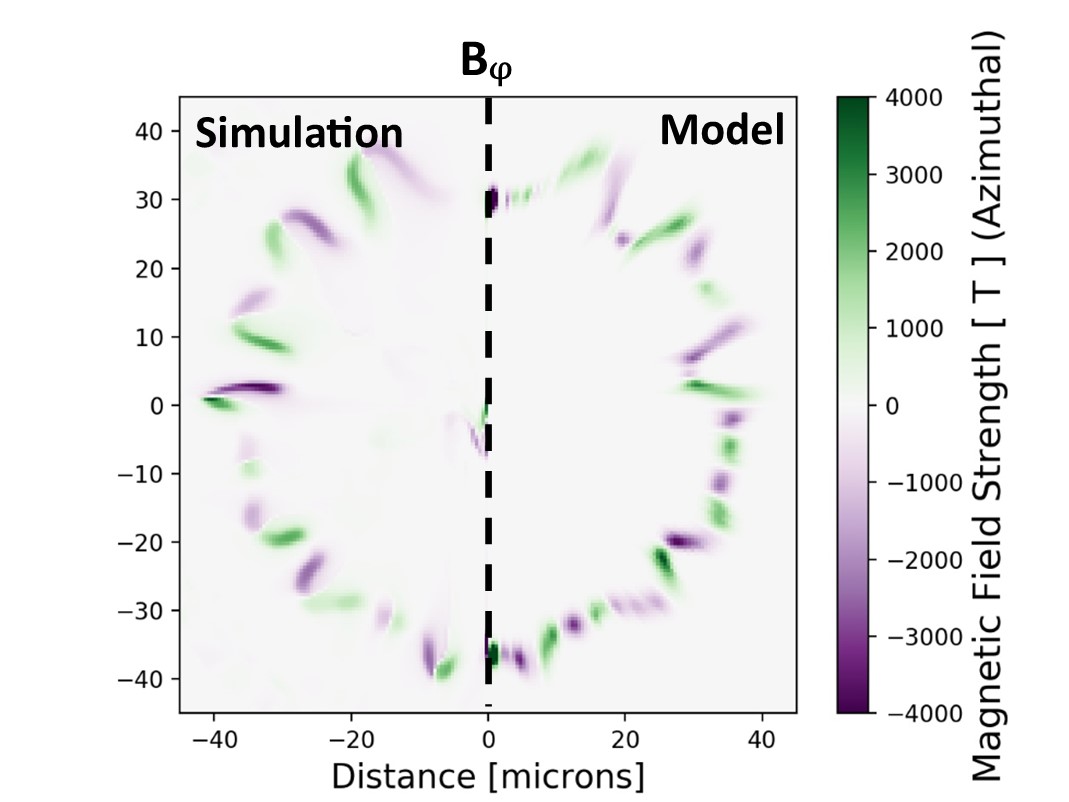}
		\caption{Left: simulated magnetic field distribution. Right: Magnetic field profile reached by taking the modeled magnetic flux distribution from figure \ref{fig:flux_vs_theta} and assuming the flux is all advected to the hot-spot edge by Nernst.}
		\label{fig:Field_Reconstructa}
	\end{subfigure}
	\begin{subfigure}[b]{0.5\textwidth}
		\centering
		\includegraphics[width=1.\textwidth]{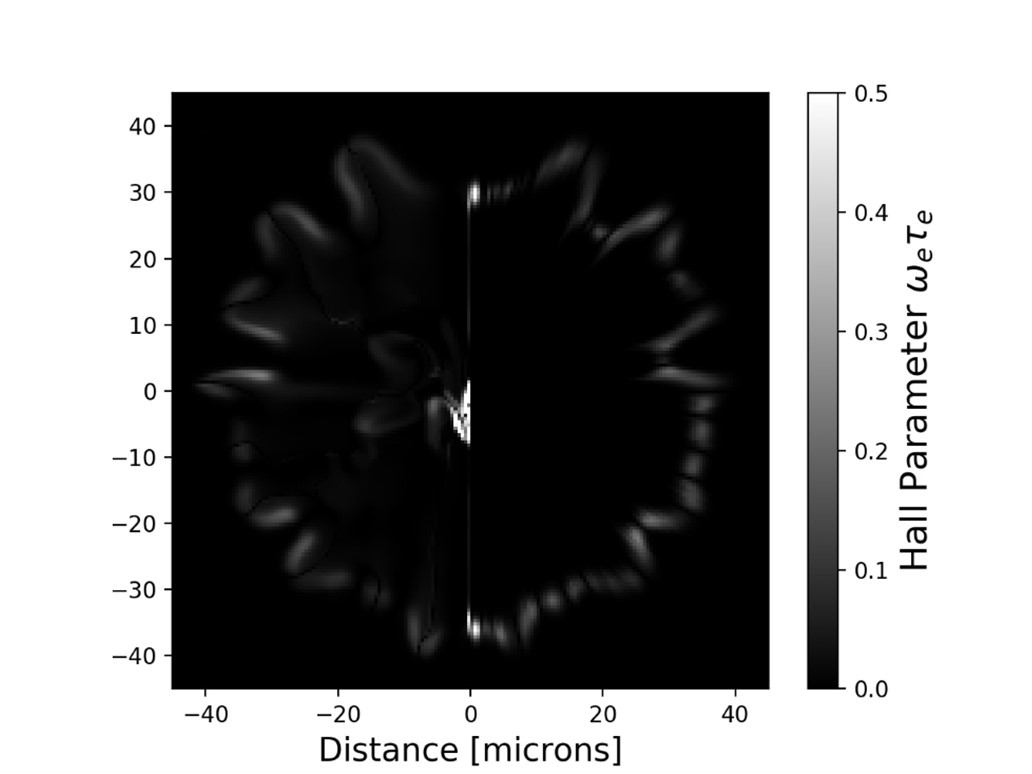}
		\caption{The electron Hall Parameter calculated by the full extended-MHD simulation (left) and predicted by the magnetic field reconstruction model (right). The high Hall Parameter in the hot-spot centre of the simulated domain originates from drive-phase magnetic flux generation.}
		\label{fig:Hall_Reconstructb}
	\end{subfigure}
	\caption{}
	\label{fig:Hall_Reconstruct}
\end{figure}

	Until now the primary focus has been on the flux generation rate. Now the focus turns to the transport of magnetic flux in the hot-spot. From equations \ref{eq:mag_trans_new} and \ref{eq:mag_trans_new_velocity} it can be seen that this transport occurs through resistive diffusion, bulk plasma advection, Nernst advection down temperature gradients and cross-gradient-Nernst advection. The Nernst term is analogous to the magnetic field advecting with the electron heat-flow within the plasma \cite{walsh2020,haines1986}, while the cross-gradient-Nernst is magnetic field moving with the Righi-Leduc heat-flow \cite{walsh2020,sadler2021}.
	
	In the radial direction the dominant transport is by bulk plasma advection and the Nernst term. As heat conducts from the hot-spot core into the surrounding shell, magnetic field is advected. However, the thermal conduction then results in ablation of shell plasma into the hot-spot, with an analogous transport of the magnetic field. In this way, the magnetic field moves with th electron energy density. Nonetheless, the Nernst term effectively confines the magnetic field to the hot-spot edge \cite{walsh2017}. 
	
	Nernst, like the heat-flow, is suppressed for larger magnetic field strengths. If the self-generated fields grow large enough, this may result in magnetic fields remaining in the hot low density hot-spot core, increasing the magnetization further. In the simulations investigated here, however, the magnetizations remain relatively low ($\omega_e \tau_e <1$) and Nernst is relatively unsuppressed.
	
	Up until now the model described by equation \ref{eq:scaling} has been shown to be effective at reproducing the magnetic flux as a function of $\theta$ (figure \ref{fig:flux_vs_theta}) in a capsule. By assuming Nernst pushes the flux to the hot-spot edge, it is possible to fully reconstruct a magnetic field profile using the model. For this end, the heat-flow divergence is calculated, and the magnetic flux is placed at the location where the most heat is being deposited. A radial smoothing is then applied on the order of the temperature scale length.
	
	Figure \ref{fig:Field_Reconstructa} compares the magnetic field strength predicted by the model to the full extended-MHD simulation at the time of peak neutron production. Again, the model compares favorably with the full simulation, with comparable peak field strengths of around $4000T$. 
	
	One discrepancy arises from the simulation including magnetic flux generated throughout the capsule implosion; flux produced during the drive-phase can be seen in the center of the simulated hot-spot. The model also does not take into account any non-radial motion of magnetic flux. Non-radial transport arises due to residual kinetic energy in the hot-spot, as well as the non-radial temperature gradients advecting the magnetic field by Nernst. For cold spikes with $\omega_e \tau_e \approx 1$ the cross-gradient-Nernst also moves the magnetic flux towards the base of the spike, which is the analogue of the Righi-Leduc heat-flow \cite{walsh2017}.
	
	Another figure of merit for the magnetic field reconstruction method can be reached by comparing the electron magnetization of the simulation with that predicted by the model. As the magnetic field is only thought to affect the plasma through magnetization of the electron population \cite{walsh2017}, it is important that the model predicts this reasonably well. Figure \ref{fig:Hall_Reconstruct} shows such a comparison. In most locations the model compares favorably, except in the hot-spot center where the drive-phase magnetic flux can magnetize the plasma and is not accounted for in the model. 
	
	As this reconstruction technique is based around a spherical co-ordinate system, the methodology fails for highly perturbed hot-spots where the hot-spot edge has multiple locations for a given $\theta,\phi$. Also, highly perturbed cases exhibit enhanced non-radial transport of the magnetic field, which is not incorporated into the model at this time. Appendix \ref{sec:model_tests} demonstrates the model breaking down for these more difficult cases.
	
	For the simulations shown here, resistive diffusion has little impact on the magnetic field distribution. Diffusion is greatest on the smallest scales, with approximate diffusion length-scales of $\frac{1}{4}\mu$m at the hot-spot edge at bang-time. Magnetic fields generated at smaller length-scales would be expected to be diffused. Equation \ref{eq:scaling} highlights how field generation is greatest around high mode perturbations. The simulations shown here are limited to $\frac{1}{2}\mu$m resolution; if shorter wavelengths are sustained in ICF hot-spots then resistive diffusion is expected to become an important process.

	\section{Conclusions}
	
	A scaling of magnetic flux generation by the Biermann Battery process has been presented and compared with 2-D extended-MHD simulations for the stagnation phase of ICF implosions. The model predicts that both higher amplitude and higher mode number perturbations generate the greatest magnetic flux. 
	
	The product of hot-spot electron temperature and density differentials can be used as a simple metric for the time-dependent generation of magnetic flux. If this product is largest at early hot-spot formation the plasma will be dominated by early magnetic flux growth, which is typically from higher mode features. Even when the high mode perturbations are ablatively stabilized around bang-time, the high mode magnetic flux is still present. The derived model can be used to compare different ICF designs, which will be the subject of future research. If, as has long been anticipated, magnetic fields are affecting the hot-spot temperature \cite{clark2016} or perturbation growth \cite{walsh2017}, the designs generating most magnetic flux should show the greatest discrepancies between experiments and rad-hydro simulations.

	By assuming Nernst deposits all of the flux at the hot-spot edge, a full magnetic field profile can be reconstructed. This can be used to estimate the impact of electron magnetization on a particular design. It may be possible to build a reduced model of thermal conductivity suppression using this model. Extended-MHD simulations are involved and cumbersome, making a simplified model appealing, particularly in 3-D simulations. As the magnetic field lines are generated along isotherms \cite{walsh2017}, the heat-flow along field lines can be neglected in these systems; the standard isotropic thermal conductivity could be lowered in accordance with the model outputs. However, the importance of Righi-Leduc heat-flow complicates the matter \cite{walsh2021a}, as inclusion of this term requires specially designed heat-flow algorithms.
	
	The presented model has limitations. Firstly, the field generation is purely by Biermann Battery, which is found to be dominant for the shell thickness perturbations applied here. For perturbation sources inducing mix, such as a fill-tube, magnetic flux generation by ionization gradients must also be included \cite{sadler2020a}. Also, the model assumes magnetic flux moves purely radially; for highly perturbed hot-spots this assumption breaks down. 
	
	As presented, the model is a post-processing tool. The magnetization of the electron population is expected to enhance perturbation growth \cite{walsh2021a}. Therefore, the use of the model on radiation hydrodynamics data is expected to under-predict the actual magnetic flux by under-estimating the $\rho R$ and $T R$ variations.

	Finally, the model can be used to understand how magnetic fields affect implosions as the designs move into the ignition regime. On the one hand, the increased hot-spot temperature will enhance magnetic flux generation. On the other hand, control of perturbation growth is required in order to reach such a regime; a return of designs to 1-D behavior would limit the impact of self-generated magnetic fields.  
	\begin{figure}
		\centering
		\begin{subfigure}[b]{0.45\textwidth}
			\centering
			\includegraphics[width=1.\textwidth]{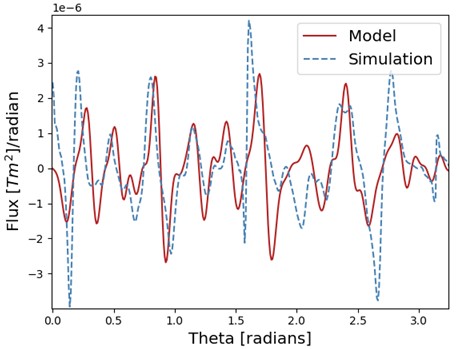}
			\caption{Comparison of magnetic flux as a function of $\theta$ between simulation and model for a large multi-mode perturbation simulation. While the peaks have moved non-radially in the simulation, the flux distribution given by the model is reasonable.}
			\label{fig:Flux_vs_theta_appendix}
		\end{subfigure}
		\begin{subfigure}[b]{0.45\textwidth}
			\centering
			\includegraphics[width=1.\textwidth]{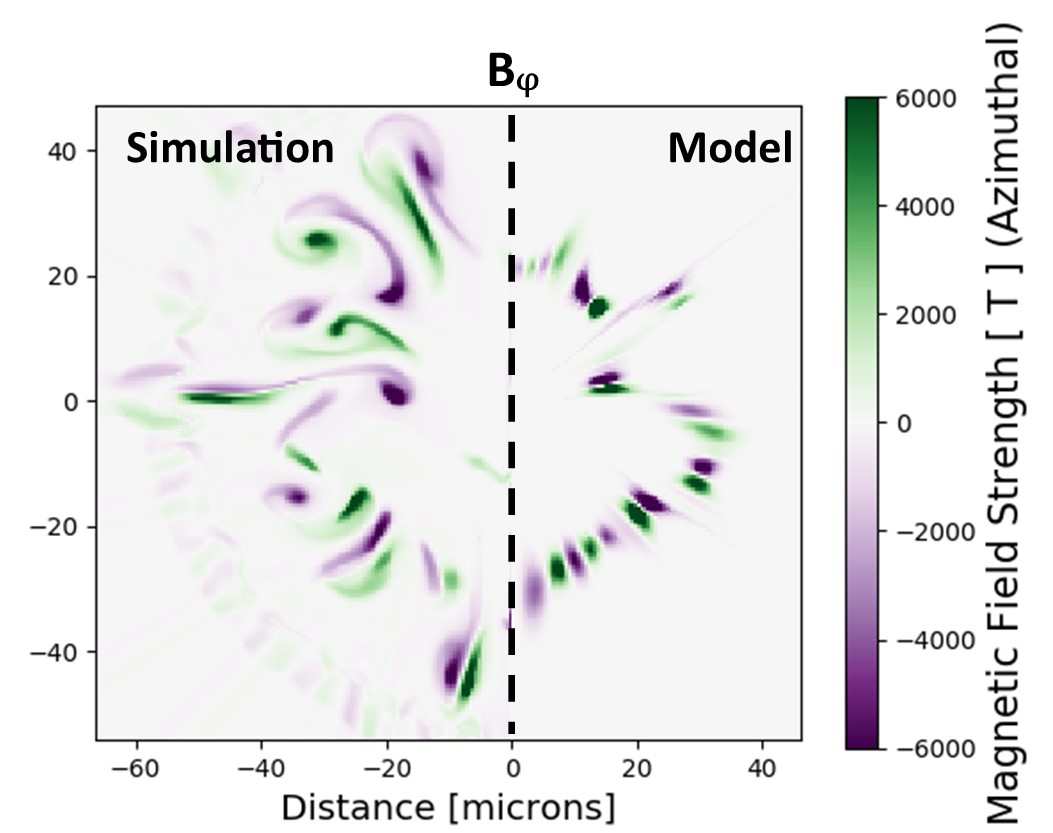}
			\caption{Comparison of magnetic field profiles given by a 2-D extended-MHD simulation (left) and by assuming all of the flux is deposited at the hot-spot edge by Nernst (right). The field reconstruction method has failed in this case, as the hot-spot edge is highly perturbed and does not have a single location in $\theta$.}
			\label{fig:Field_Reconstruct_appendix}
		\end{subfigure}
		\caption{Magnetic flux generation model (a) and subsequent field reconstruction (b) for a simulation with a large $40nm$ initial perturbation amplitude. Both figures are for bang-time (8.5ns).}
		\label{fig:LargeAmp}
	\end{figure}
\begin{figure*}
	\centering
	\includegraphics[scale=0.5]{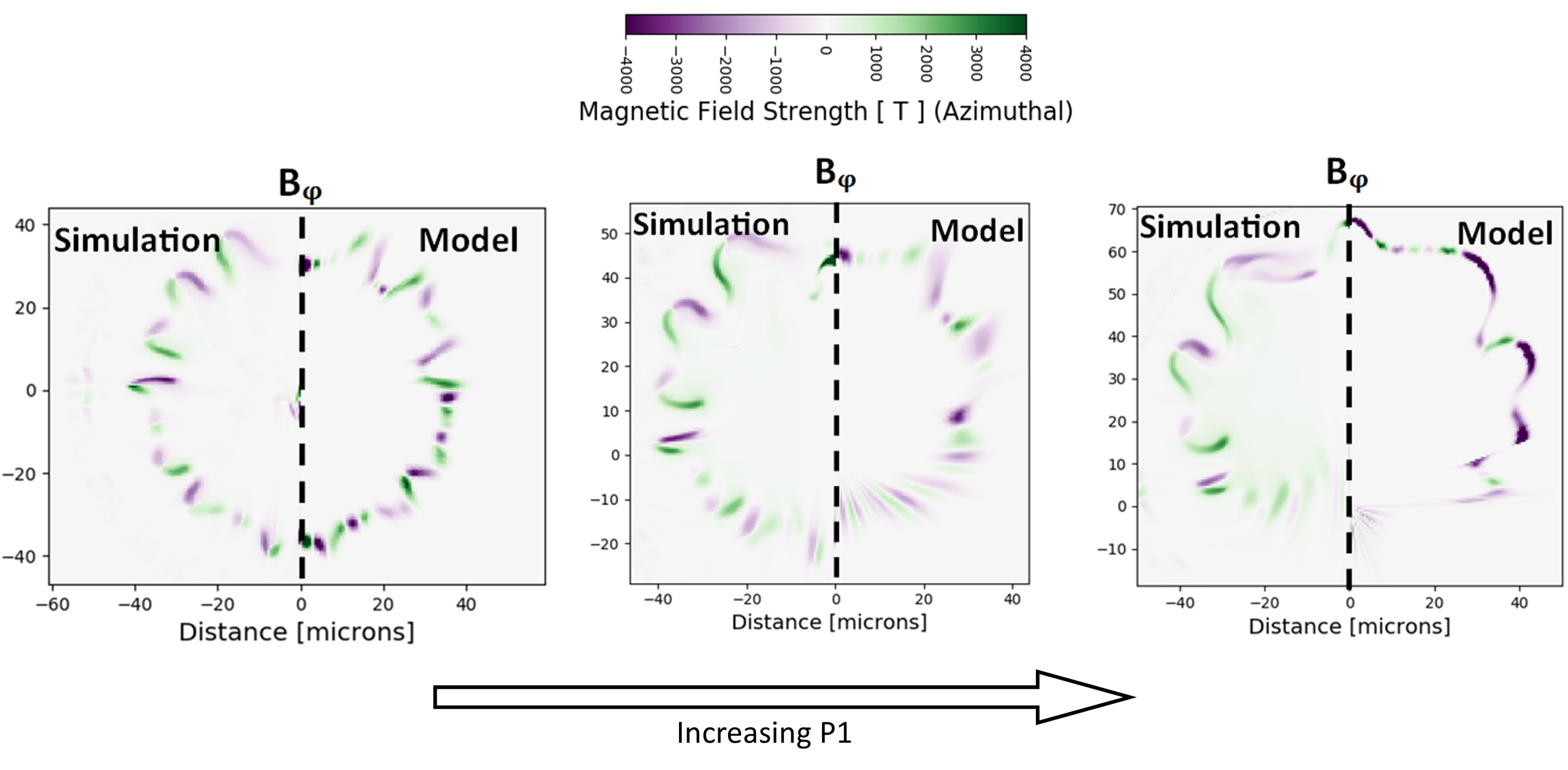}\caption{ Magnetic field reconstruction for a range of P1 amplitudes. Left: no P1 applied. Center: 1\% P1 drive. Right: 2\% P1 drive. As the magnetic flux model uses spherical co-ordinates, it is not well adapted to incorporate mode 1 asymmetries.  \label{fig:P1}  }
\end{figure*}
	
	%\section{Appendix A: Biermann Catastrophe \label{sec:catastrophe}}
	
	\section{Appendix A: Stress-testing the Magnetic Flux Model \label{sec:model_tests}}

	Naturally, the efficacy of a model formulated around polar co-ordinates has its limits. Here the model is compared to simulations for two stress tests. First, large amplitude perturbations that deform the capsule far from spherical by bang-time. Then, an added P1 drive asymmetry that moves the capsule center is shown.
	
	Figure \ref{fig:Flux_vs_theta_appendix} compares the simulation to the first part of the model, where a magnetic flux is obtained as a function of $\theta$. This simulation is the same as that used in figure \ref{fig:flux_vs_theta}, but with all perturbation amplitudes increased by a factor of 4, resulting in a highly perturbed hot-spot. The larger perturbation results in greater non-radial magnetic transport, which is not included in the theoretical model. This can be seen by looking at how far the peaks have shifted in $\theta$ in figure \ref{fig:Flux_vs_theta_appendix} compared with figure \ref{fig:flux_vs_theta}. Nonetheless, the flux distribution gives a reasonable estimate. 
	
	Figure \ref{fig:Field_Reconstruct_appendix} then shows how the model breaks down once the magnetic flux is placed radially. As the hot-spot is highly perturbed, the hot-spot radius is multi-valued for a given $\theta$. Clearly, care should be taken when using this model on highly perturbed simulations. 
	
	The spherical co-ordinate system used for the model also requires an assumed hot-spot center. This becomes problematic when a mode 1 asymmetry is present \cite{spears2014}, shifting the hot-spot center with time. To combat this, the model has been adapted to track the hot-spot center of mass, each timestep assuming that the whole hot-spot, including the magnetic flux, has shifted by a uniform distance. To test this, again the simulation used in figure \ref{fig:Hall_Reconstruct} is re-run, but this time with applied P1 drive asymmetries of 1\% and 2\% in radiative energy. The P1 is applied as a constant throughout the prescribed X-ray drive. The 1\% P1 results in a hot-spot center of mass shifted by approximately $20 \mu m$, while the 2\% P1 results in a $30 \mu m$ offset.
	
	Figure \ref{fig:P1} shows how the model fares as the mode 1 increases. The total flux remains well estimated, but the spatial distribution of the reconstruction increases in error with larger drive asymmetry. In particular, the temperature length-scale over which to spread the magnetic field radially is underestimated.

	\section*{Acknowledgements}

	This work was performed under the auspices of the U.S. Department of Energy by Lawrence Livermore National Laboratory under Contract DE-AC52-07NA27344.

	This document was prepared as an account of work sponsored by an agency of the United States government. Neither the United States government nor Lawrence Livermore National Security, LLC, nor any of their employees makes any warranty, expressed or implied, or assumes any legal liability or responsibility for the accuracy, completeness, or usefulness of any information, apparatus, product, or process disclosed, or represents that its use would not infringe privately owned rights. Reference herein to any specific commercial product, process, or service by trade name, trademark, manufacturer, or otherwise does not necessarily constitute or imply its endorsement, recommendation, or favoring by the United States government or Lawrence Livermore National Security, LLC. The views and opinions of authors expressed herein do not necessarily state or reflect those of the United States government or Lawrence Livermore National Security, LLC, and shall not be used for advertising or product endorsement purposes.

	\section*{Data Availability}
	The data that support the findings of this study are available from the corresponding author upon reasonable request.
		
	\section*{References}
	
	\bibliographystyle{plainnat}
	\ifdefined\DeclarePrefChars\DeclarePrefChars{'’-}\else\fi

\end{document}